\documentclass[a4paper,11pt]{article}
\usepackage{pos}

\usepackage{float}
\usepackage{graphicx}

\newcommand{\m}{m_{\text{P}}}

\begin{document}

\title{\boldmath
  Explaining the Hubble tension and dark energy from $\alpha$-attractors}
\ShortTitle{Non-oscillatory EDE}
\author{\ Lucy Brissenden}

\author*{Konstantinos Dimopoulos}

\author{Samuel S\'anchez L\'opez}

\affiliation{Consortium for Fundamental Physics, Lancaster University,\\
  Lancaster LA1 4YB, UK}


\emailAdd{l.brissenden@lancaster.ac.uk}
\emailAdd{k.dimopoulos1@lancaster.ac.uk}
\emailAdd{s.sanchezlopez@lancaster.ac.uk}

\abstract{A compelling unified model of dark energy and early dark energy (EDE)
  is presented, using a scalar field with an exponential runaway potential, in
  the context of alpha-attractors. The field is originally trapped at an
  enhanced symmetry point, subsequently thaws to become successful EDE and
  eventually slow-rolls to become dark energy. EDE ameliorates the observed
  Hubble tension.}

\FullConference{%
  Corfu Summer Institute 2022 "School and Workshops on Elementary Particle Physics and Gravity",\\
  28 August - 1 October, 2022\\
  Corfu, Greece}


\maketitle

\section{Introduction}

Since the arrival of high precision cosmological observations, a standard model
of cosmology, called the concordance model, has been constructed. In short
called $\Lambda$CDM, this concordance model recently has starting to suffer
from a number of challenges, the most important of which is arguably the
Hubble tension.

In a nutshell, the Hubble tension amounts to a disagreement in the estimate of
the Hubble constant $H_0$ as inferred by early (mainly CMB) and late (mainly
SNe) observations. Indeed, CMB observations from the Planck satellite
\cite{Planck2018} suggest
\begin{equation}
H_0= 67.44\pm0.58 \ \textnormal{km} \ \textnormal{s}^{-1}\textnormal{Mpc}^{-1},
\end{equation}    
while a distance scale measurement using Cepheid-SN-Ia data 
from the SH0ES collaboration \cite{sh0es2021}
results in
\begin{equation}
H_{0} = 73.04\pm1.04 \ \textnormal{km} \ \textnormal{s}^{-1}\textnormal{Mpc}^{-1}.
\end{equation}

This is a $5\sigma$ tension (4.56-6.36$\sigma$). The most compelling proposal to
overcome this problem is early dark energy (EDE). 

\section{Early Dark Energy}
\label{subsec:ede}

EDE amounts to momentary importance of dark energy near matter-radiation
equality. Investigated first by
Refs.~\cite{stringaxiverse,howearly,Calabrese_2011,Doran_2006},
this proposal does not necessarily consider that EDE is the same dark energy
substance which is responsible for the accelerated expansion at present (for a recent review see Ref.~\cite{EDEreview}). 

How does EDE manage to increase the value of $H_0$ as inferred from CMB observations? Even
though CMB and BAO observations tightly constrain the cosmological parameters, 
they constrain the combination $H(z)r_s$, where $H(z)$ is the Hubble parameter 
as a function of redshift $z$,
and $r_s$ is the comoving sound horizon at decoupling, given by 
\begin{equation}
    r_s = \int^{\infty}_{z_{\rm dec}}\frac{c_{s}(z)}{H(z)}dz,
    \label{rs}
\end{equation}
where $c_{s}(z)$ is the sound speed. An additional amount of dark energy in the
Universe increases the total density, which in turn increases the Hubble
parameter because of the Friedmann equation
\mbox{$H^2=(\rho_B+\rho_{_{\rm EDE}})/3m_{\m}^2$}, where $\rho_B$ is the density of
the radiation and matter background. EDE amounts to a brief increase of $H(z)$
before decoupling, which lowers the value of the sound horizon in
Eq.~(\ref{rs}). Thus, EDE manages to simultaneously lower the value of $r_s$
and increase $H_{0}$ without violating CMB observations.

The fractional energy density required for EDE to work is about 10\%, 
\mbox{$f_{_{\rm EDE}}=0.10\pm0.02$} at redshift \mbox{$z_c=4070^{+400}_{-840}$}.
Therefore, the EDE proposal amounts to as injection of energy at around the
time of matter-radiation equality (\mbox{$z_c\simeq z_{\rm eq}\simeq 3600$}),
which then decays away faster than the background radiation, such that it
becomes negligible at the time of last scattering, before it can be detected in
the CMB \cite{howearly}.

The original proposal in Ref.~\cite{stringaxiverse} suggested that the EDE
was an axion scalar field \mbox{$\phi=\theta f$} with potential
\mbox{$V(\theta)=m^2f^2(1-\cos\theta)^n$} with \mbox{$n>2$}. The authors of Ref.~\cite{stringaxiverse} found that
the fractional energy density must be \mbox{$f_{_{\rm EDE}}=0.08\pm0.04$}, which
results in \mbox{$H_0=70.0\pm1.5\,$km$\,{\rm s}^{-1}{\rm Mpc}^{-1}$}.
After thawing, the EDE field oscillates around its vacuum expectation value (VEV)
with average barotropic parameter \mbox{$w=\frac{n-1}{n+1}$}. To redshift
faster than radiation, it is needed that \mbox{$w>\frac13$}, which implies that the minimum is of
order higher than quartic. Note that the density of the oscillating EDE
redshifts as $a^{-6n/(n+1)}$, which reduces to $a^{-6}$ (free-fall) in the limit
\mbox{$n\gg 1$}. The situation is similar to many other EDE models, where
typically the EDE scalar field oscillates around its VEV in a high order
potential (see however, Ref.~\cite{referee}). 
In contrast, in our model presented below, the EDE scalar field
experiences a period of kinetic domination, where the field is in
non-oscillatory free-fall and its density decreases as $\propto a^{-6}$.

\section{\boldmath $\alpha$-attractors}
\label{subsec:alphaattractors}

Our model unifies EDE with late dark energy in the context of
$\alpha$-attractors. Ref.~\cite{alphaattractorsede1} is an earlier attempt for
such unification in the same theoretical context. However, that proposal
is also of oscillatory EDE.

$\alpha$-attractors 
appear naturally in conformal field theory or supergravity theories \cite{lindealphaattractors, Kallosh:2013hoa,  Ferrara:2013rsa, Kallosh:2014rga}.
The scalar field has a non-canonical kinetic term, featuring two poles, which
the field cannot cross. The field can be canonically normalised via a
field redefinition. Then, the finite poles for the non-canonical field are
transposed to infinity for the canonical one. As a
result, the scalar potential is ``stretched'' near the poles, featuring two
plateau regions, which have been used for modelling inflation, see \textit{e.g.}
Refs. \cite{Alho:2017opd, Odintsov:2016vzz, Braglia:2022phb, Kallosh:2022ggf, Achucarro:2017ing, Iarygina:2020dwe, DimopoulosWaterfall}  or quintessence \cite{Akrami:2017cir},
or both, in the context of quintessential inflation \cite{Dimopoulos:2017tud, Dimopoulos:2017zvq, Akrami:2017cir}.

The Lagrangian density features two poles at $\varphi=\pm\sqrt{6\alpha}\,\m$ and has
the form
\begin{equation}
  \mathcal{L} = \frac{-\frac{1}{2}(\partial\varphi)^{2}}%
          {\left(1-\frac{\varphi^{2}}{6\alpha \,\m^{2}}\right)^{2}}
          - V(\varphi)\,,
    \label{L0}
\end{equation}
where $\varphi$ is the non-canonical scalar field and
\mbox{$(\partial \varphi)^2\equiv g^{\mu\nu}\partial_{\mu}\varphi\,\partial_{\nu}\varphi$}. Redefining $\varphi$ in terms of the canonical scalar field $\phi$, we
have
\begin{equation}
    \text{d}\phi=\frac{\text{d}\varphi}{1-\frac{\varphi^2}{6\alpha \m^2}}\quad \Rightarrow\quad\varphi = \m\sqrt{6\alpha}\,\tanh{\left(\frac{\phi}{\sqrt{6\alpha}\,\m}\right)}\,.
    \label{phivarphi}
\end{equation}
The poles $\varphi=\pm \sqrt{6\alpha}\,\m$ are transposed to infinity and the Lagrangian density now reads

\begin{equation}
    \mathcal{L} = -\frac{1}{2}(\partial\phi)^{2} - V(\phi).
    \label{L}
\end{equation}

\section{The Model}
\label{sec:model}


In contrast to most EDE literature, we investigate non-oscillating EDE. Thus, we require the scalar potential to be steep enough, such that, after equality of matter and radiation, the EDE scalar field becomes dominated by its kinetic energy density and engages in "free-fall" roll. Therefore, we study the following toy-model.

Consider a potential of the form
\begin{equation}
  V(\varphi) = V_X\exp(-\lambda e^{\kappa\varphi/\m}),
      \label{Vvarphi}
\end{equation}
where $\alpha, \kappa, \lambda$ are dimensionless model parameters, $V_X$ is a
constant energy density scale and $\varphi$ is the non-canonical scalar field
with kinetic poles given by the typical $\alpha$-attractors form
with the Lagrangian density in Eq.~(\ref{L0}).

To assist our intuition, we switch to the canonically normalised (canonical)
scalar field $\phi$, using the transformation in Eq.~\eqref{phivarphi}.
The Lagrangian density is then given by Eq.~\eqref{L}, where the scalar
potential is
\begin{equation}
    V(\phi) = \exp(\lambda e^{\kappa\sqrt{6\alpha}})V_{\Lambda}\exp[-\lambda e^{\kappa\sqrt{6\alpha}\tanh(\phi/\sqrt{6\alpha}\,\m)}]\,,
    \label{eqn:currentpotential}
\end{equation}
where $V_\Lambda$ is the vacuum density at present related to the model
parameters as
\begin{equation}
V_\Lambda\equiv \exp(-\lambda e^{\kappa\sqrt{6\alpha}})V_X\,.
\label{VL}
\end{equation}
Note that the model parameter is $V_X$ and not $V_\Lambda$, the latter being
generated by $V_X$ and the remaining model parameters as shown above.

\section{Analytic study}

We are interested in two limits for the
potential: matter-radiation equality and the present time.
At matter-radiation equality, we consider \mbox{$\phi\rightarrow 0$}
 ($\varphi\rightarrow 0$). In this limit, we have
\begin{equation}
    V_{\textnormal{eq}} \simeq \exp[\lambda(e^{\kappa\sqrt{6\alpha}}-1)]V_{\Lambda}\exp(-\kappa\lambda\,\phi_{\textnormal{eq}}/\m)\,,
    \label{eqn:lowphiapprox}
\end{equation}
where the subscript `eq' denotes 
the time of matter-radiation equality. 
It is assumed that the field was originally frozen there and at the time of
equality in unfreezes (thaws). We discuss and justify this assumption in
Sec.~\ref{sec:ESP}.

After thawing the field soon rolls towards large values. Today, we consider
 \mbox{$\phi\rightarrow+\infty$} ($\varphi\rightarrow+\sqrt{6\alpha}\,\m$).
 The potential in this limit is
\begin{equation}
    V_{0} \simeq V_{\Lambda}\left[1+2\kappa\lambda e^{\kappa\sqrt{6\alpha}}\sqrt{6\alpha}\,\exp\left(-\frac{2\phi_{0}}{\sqrt{6\alpha}\,\m}\right)\right],
    \label{eqn:highphiapprox}
\end{equation}
where the subscript `0' denotes the present time. Note that, in this limit, the
potential approaches $V_\Lambda$, which corresponds to positive vacuum density
with $w=-1$, as in $\Lambda$CDM.

The above approximations describe well the scalar potential near equality and
the present time. As explained below, between these regions, the scalar field
free-falls and becomes oblivious of the scalar potential.

Let us investigate the evolution of the EDE field.
Originally the field is frozen at zero (see Sec.~\ref{sec:ESP}).
Its energy density is such that it remains frozen there until equality, when it
thaws following the appropriate exponential attractor, since $V_{\rm eq}$ in
Eq.~\eqref{eqn:lowphiapprox} is approximately exponential
\cite{Copeland:1997et}.

For convenience, we assume 
this is the subdominant attractor, which requires that the strength of the
exponential is \cite{Copeland:2006wr,kostasbook}
\begin{equation}
  \kappa\lambda>\sqrt 3\,.
  \label{z}
  \end{equation}
The subdominant exponential attractor is called the scaling attractor. In
the scaling attractor the energy density of the rolling scalar field mimics the
dominant background energy density. Thus, the fractional energy density
of the field is constant, given by the value
\cite{Copeland:1997et,Copeland:2006wr,kostasbook}
\begin{equation}
f_{_{\rm EDE}}
\simeq
\frac{3}{(\kappa\lambda)^2}<1
\label{Omegaphieq}
\end{equation}
This provides an estimate of the moment when the originally frozen scalar field,
unfreezes and begins rolling down its potential. Before unfreezing
$f_{_{\rm EDE}}$ is growing, because the background
density decreases with the expansion of the Universe, until $f_{_{\rm EDE}}$
obtains the above value.

However, after unfreezing, the field soon experiences the full exp(exp) steeper
than exponential potential so, it does not follow the subdominant attractor any
more but it is dominated by its kinetic energy density only
(it free-falls). Then, its density scales as
\mbox{$\rho_\phi\simeq\frac12\dot\phi^2\propto a^{-6}$}, until it refreezes at a
larger value $\phi_0$. This value is estimated as follows.

In free-fall, the equation of motion is reduced to
\mbox{$\ddot\phi+3H\dot\phi\simeq0$}, where $H=2/3t$ after equality.
The solution is
\begin{equation}
\phi(t)=\phi_{\rm eq}+\frac{C}{t_{\rm eq}}\left(1-\frac{t_{\rm eq}}{t}\right)\,,
\label{phit}
\end{equation}
where $C$ is an integration constant. From the above, it is straightforward
to find that \mbox{$\dot\phi=Ct^{-2}$}. Thus, at equality we have
\begin{eqnarray}
  &&  
  f_{_{\rm EDE}}=\left.\frac{\rho_\phi}{\rho}\right|_{\rm eq}=
  \frac{\frac12 C^2 t_{\rm eq}^{-4}}{\frac43(\frac{\m}{t_{\rm eq}})^2}=
  \frac38\frac{C^2}{(\m t_{\rm eq})^2}\nonumber\\
\Rightarrow &&
C=\sqrt{\mbox{$\frac83$}
f_{_{\rm EDE}}}\,\m\,t_{\rm eq}=
\frac{\sqrt 8}{\kappa\lambda}\,\m\,t_{\rm eq}\;,
\label{C}
\end{eqnarray}
where we used Eq.~(\ref{Omegaphieq}), \mbox{$\rho_\phi\simeq\frac12\dot\phi^2$}
and that
\mbox{$\rho=1/6\pi Gt^2=\frac43(m_P/t)^2$}. Therefore, the field freezes at the
value
\begin{equation}
  \phi_0=\phi_{\rm eq}+C/t_{\rm eq}=\phi_{\rm eq}+
  \frac{\sqrt 8}{\kappa\lambda}\,\m\;,
\label{phi0}
\end{equation}
where we considered that \mbox{$t_{\rm eq}\ll t_{\rm freeze}<t_0$}\ .

Using that \mbox{$t_{\rm eq}\sim 10^4\,$y} and \mbox{$t_0\sim 10^{10}\,$y}, we can estimate
\begin{equation}
  \frac{V_{\rm eq}}{V_0}\simeq\frac{
    f_{_{\rm EDE}}\rho_{\rm eq}}{0.7\,\rho_0}
  \simeq\frac{30}{7(\kappa\lambda)^2}
  \left(\frac{t_0}{t_{\rm eq}}\right)^2\simeq
  \frac{3}{7(\kappa\lambda)^2}\times 10^{13}\,.
  \label{Vratio}
\end{equation}
Now, from Eqs.~(\ref{eqn:lowphiapprox}) and (\ref{eqn:highphiapprox}) we find
\begin{equation}
 \frac{V_{\rm eq}}{V_0}\simeq
 \frac{e^{\lambda(e^{\kappa\sqrt{6\alpha}}-1)}\exp(-\kappa\lambda\,\phi_{\rm eq}/\m)}{1+
   2\kappa\lambda\,e^{\kappa\sqrt{6\alpha}}\sqrt{6\alpha}
   \exp(-2\phi_0/\sqrt{6\alpha}\,\m)}\,.
\end{equation}
Considering that \mbox{$\phi_{\rm eq}\simeq 0$} and Eq.~(\ref{phi0}), the above
can be written as
\begin{equation}
  \frac{V_{\rm eq}}{V_0}\simeq
  \frac{e^{\lambda(e^{\kappa\sqrt{6\alpha}}-1)}}{1+ 2\kappa\lambda\,
    e^{\kappa\sqrt{6\alpha}}\sqrt{6\alpha}\,e^{-2\sqrt 8/\kappa\lambda\sqrt{6\alpha}}}\,.
  \label{Vratio+}
\end{equation}

Taking \mbox{$
  f_{_{\rm EDE}}\simeq 0.1$} as required by EDE, Eq.~(\ref{Omegaphieq}) suggests
\begin{equation}
  \kappa\lambda\simeq\sqrt{30}\,.
  \label{30}
\end{equation}
Combining this with Eq.~(\ref{Vratio}) we obtain
\begin{equation}
  e^{\frac{\sqrt{30}}{\kappa}(e^{\kappa\sqrt{6\alpha}}-1)}\sim 10^{12}/7\,,
\label{master}
\end{equation}  
where we have ignored the 2nd term in the denominator of the right-hand-side of
Eq.~(\ref{Vratio+}).

From the above we see that, $\kappa$ is large when $\alpha$ is small. Taking, as an example,
\mbox{$\alpha=0.01$} we obtain \mbox{$\kappa\simeq 18$} and
\mbox{$\lambda\simeq 0.30$} (from Eq.~(\ref{30})). With these values,
the second term in the denominator of the right-hand-side of
Eq.~(\ref{Vratio+}) is of order unity and not expected to 
significantly influence our results.

For the selected values,
Eq.~(\ref{phi0}) suggests that the total excursion of the field is
\begin{equation}
  \Delta\phi=\phi_0-\phi_{\rm eq}=\frac{\sqrt 8}{\kappa\lambda}\,m_P
  \simeq 0.5\,m_P\;,
  \label{Dphi}
\end{equation}
i.e. it is sub-Planckian. A sub-Planckian excursion of the field implies that
5th force considerations are suppressed.

\section{Numerical investigation}

We have thoroughly analysed this model in Ref.~\cite{ours}. Here, we will present our main results.

We have aimed to obtain a value of $H_0$ in the window
\begin{equation}
72\leq\frac{H_0}{\rm km\;sec^{-1}\,Mpc^{-1}}\leq 74\,.
\label{H0range}
\end{equation}
With this requirement, the parameter space arrived at for our model parameters
is
\begin{eqnarray}
  && 0<\alpha<0.00071\nonumber\\
  && 0<\kappa<700\nonumber\\
  && 0<\lambda<0.027\;,\\
\nonumber
  \label{param}
\end{eqnarray}
with \mbox{$V_\Lambda=10^{-120.068}\,\m^4$}. We see that the above numbers are
reasonable. In particular, the value of \mbox{$\kappa\sim 10^2$} implies that
the mass scale suppressing the exponent in our model in Eq.~\eqref{Vvarphi} is
near the scale of grand unification \mbox{$\m/\kappa\sim 10^{16}\,$GeV}, which
is a rather natural scale. 

In the above ranges, we find that \mbox{$0.015<f_{_{\rm EDE}}<0.107$} at equality,
while it becomes lower than $10^{-3}$ by decoupling; the time the CMB radiation
is emitted. The barotropic parameter of dark energy at present is
\mbox{$w_\phi=-1.000$} with negligible running (less than $10^{-11}$), which is
indistinguishable from $\Lambda$CDM.

One important finding is that the condition in Eq.~\eqref{z}, \mbox{$\kappa\lambda>\sqrt 3$}
assumed in the previous section, is not valid. However,
this was chosen only for convenience, as explained before Eq.~\eqref{z}. If the condition is
violated then the thawing EDE does not follow the scaling exponential attractor
but the dominant exponential attractor instead. In both cases however, once the
EDE field rolls away from zero, it starts experiencing the full exp(exp)
potential and goes into free-fall, as discussed. Thus, the qualitative
behaviour is the same, as also demonstrated by our numerical results shown below.

\begin{figure}[h]
\centering
\includegraphics[width=.8\columnwidth]{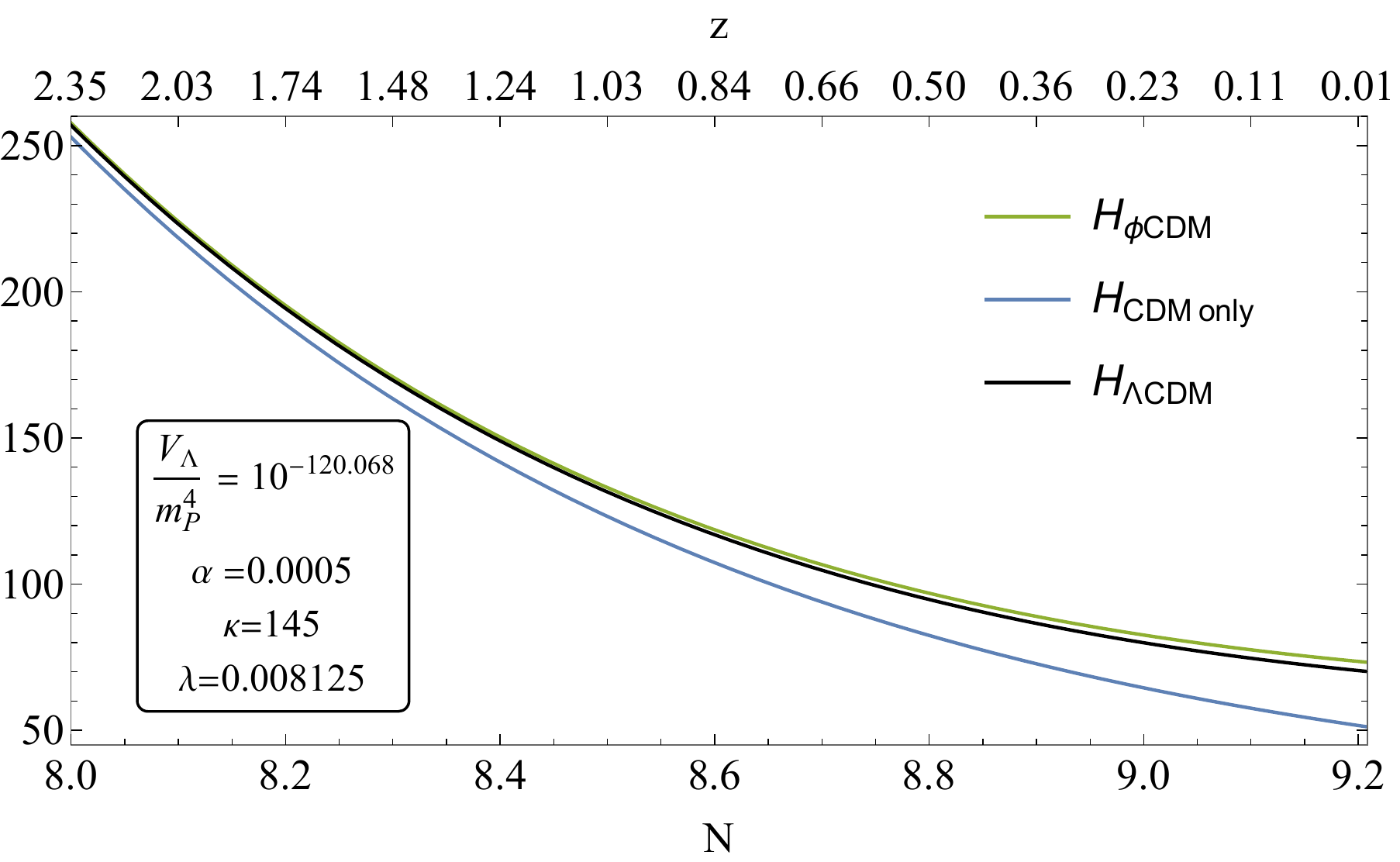}
\caption{The Hubble parameter (in units of $ \textnormal{km} \ \textnormal{s}^{-1}\textnormal{Mpc}^{-1}$) of the Universe in our model (green), a classical $\Lambda$CDM simulation (black), and one with only matter and radiation (blue), as a function of the redshift (top) and the e-folds (bottom) elapsed since the beginning of the simulation. It is evident that our model corresponds to a larger value of $H(z)$ than $\Lambda$CDM, as desired.}
\label{fig:hubble}
\end{figure}

As a concrete example, we choose the following values for the model parameters
\begin{eqnarray}
  && \alpha=0.0005\nonumber\\
  && \kappa=145\nonumber\\
  && \lambda=0.008125\,.\\
  \nonumber
  \label{concrete}
\end{eqnarray}
The above suggest that \mbox{$\kappa\lambda=1.178<\sqrt 3$}. The value of the
Hubble constant obtained in this case is
\begin{equation}
H_0=73.27\;{\rm km}\;{\rm sec}^{-1}\,{\rm Mpc}^{-1},
\label{H0concrete}
\end{equation}
which evidently is well in agreement with the SH$_0$ES observations. Comparison of
the Hubble parameter in this scenario with the one in $\Lambda$CDM is shown in
Fig.~\ref{fig:hubble}.

\begin{figure}[h]
\centering

\mbox{\hspace{1cm}}

\includegraphics[width=.8\columnwidth]{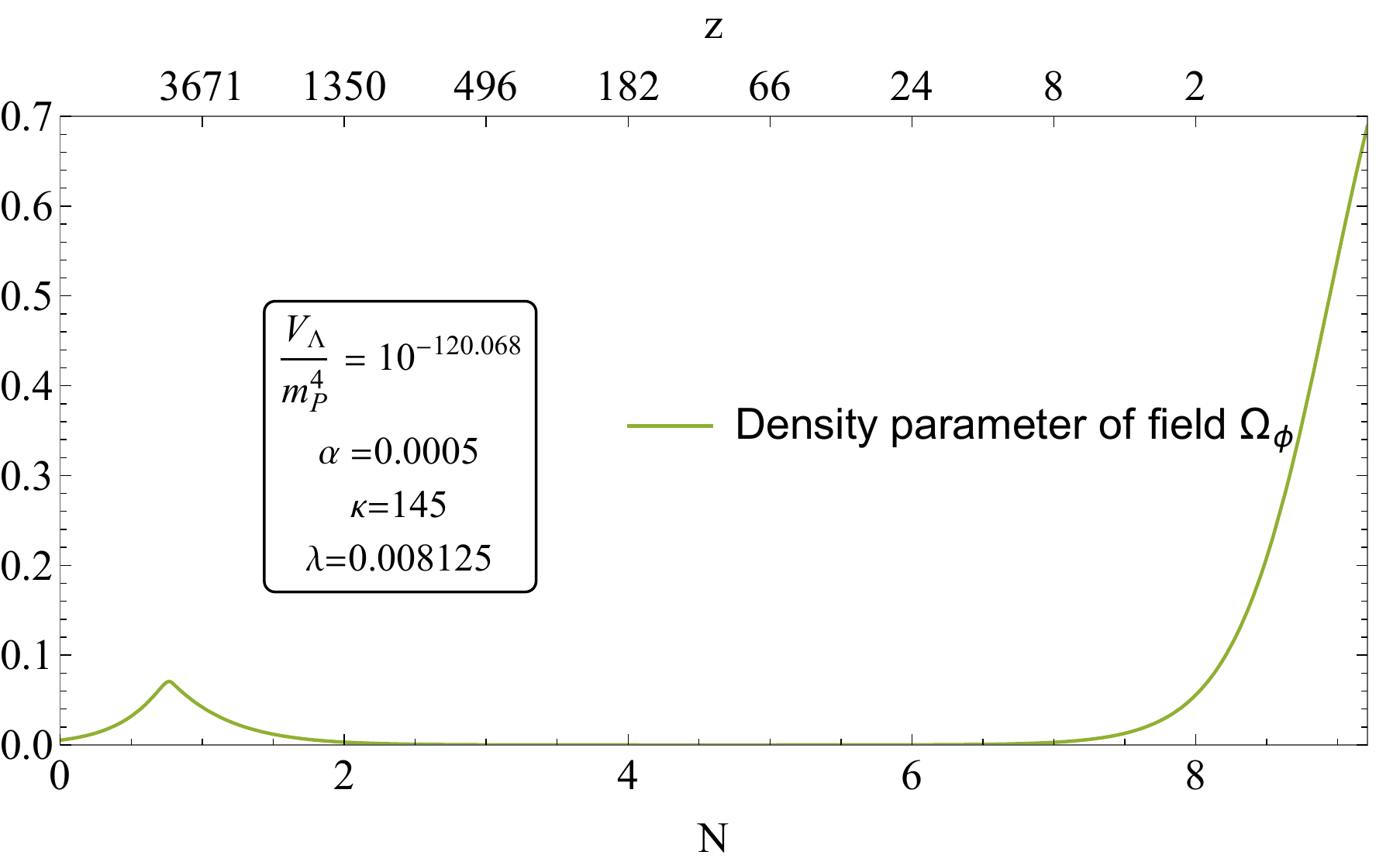}

\caption{The density parameter of the scalar field $\Omega_\phi$ as a function of the redshift (top) and e-folds (bottom) elapsed since the beginning of the simulation. As shown, at equality, there is a bump with \mbox{$f_{_{\rm EDE}}=\Omega_\phi ( z_{\rm eq} ) $} with \mbox{$f_{_{\rm EDE}}\simeq 0.08$}.}
\label{fig:fEDE}
\end{figure}

The behaviour of the fractional energy density $f_{_{\rm EDE}}$, which is
identified with the EDE density parameter $\Omega_\phi(z)$ is shown in
Fig.~\ref{fig:fEDE}. It is evident that, for this example,
\mbox{$f_{_{\rm EDE}}=\Omega_\phi(z_{\rm eq})\simeq 0.08$}.

In view of Eq.~\eqref{VL}, we find \mbox{$\log(V_X/V_\Lambda)=9.926$}. Thus,
our model parameter \mbox{$V_X=10^{-110.142}\,\m^4$} is fine-tuned at the same
level (slightly less) than $V_\Lambda$ in $\Lambda$CDM. However,
it has to be stressed that,
in contrast to $\Lambda$CDM, our proposal addresses simultaneously two cosmological problems;
not only late dark energy but also the Hubble tension.

\begin{figure}[h]
\centering
\includegraphics[width=.8\columnwidth]{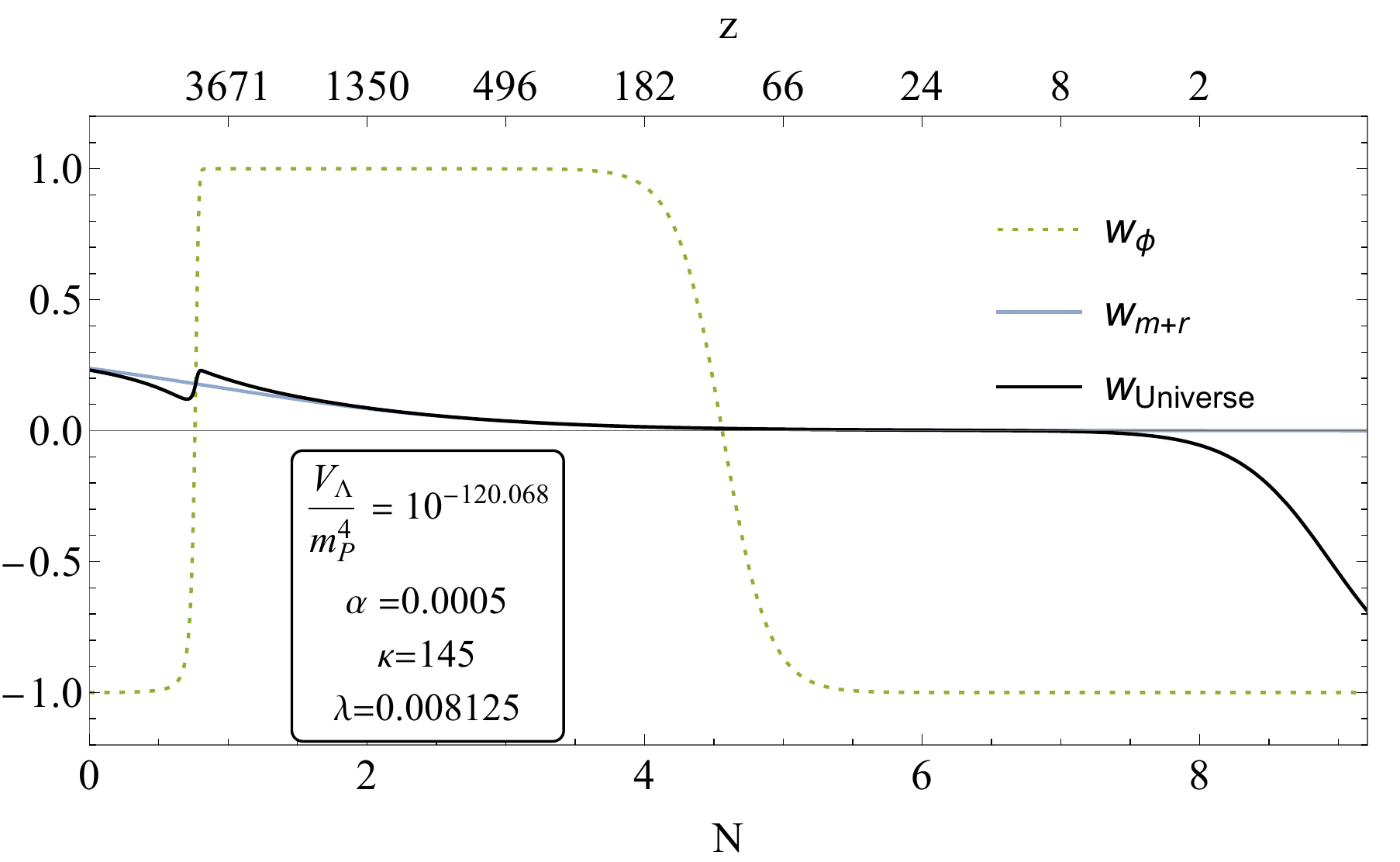}
\caption{Barotropic parameter of the scalar field (dotted green), of the background perfect fluid (full blue) and of the sum of both components (full black). It is evident that, after unfreezing, the EDE scalar field is in free-fall, with $w_\phi=1$, until it refreezes again.}
\label{fig:barotropic}
\end{figure}

\begin{figure}[h]
\centering
\includegraphics[width=.8\columnwidth]{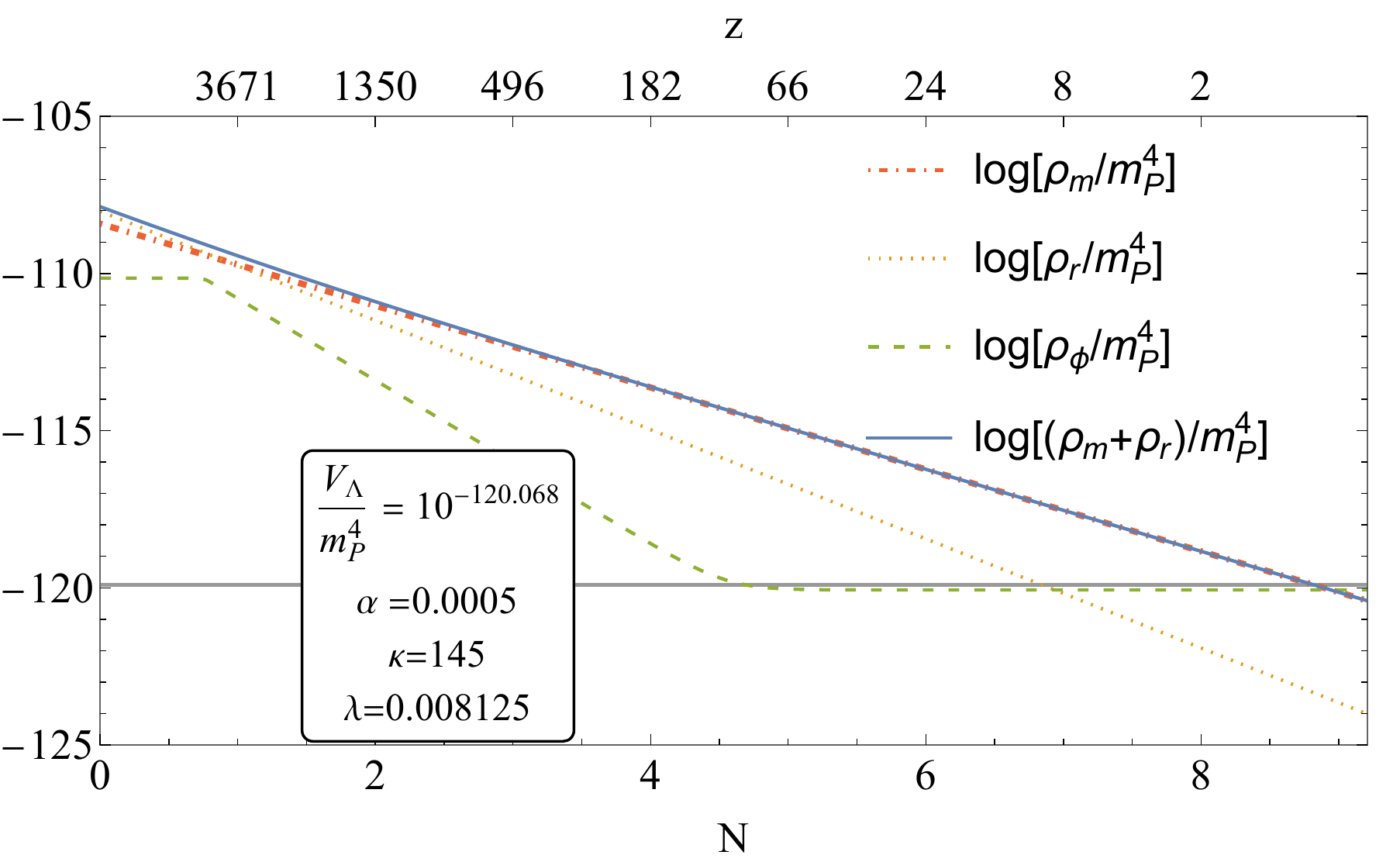}
\caption{The logarithmic densities of matter (dot-dashed red), radiation (dotted orange), the sum of both (solid blue) and the EDE scalar field (dashed green), as a function of the redshift (top) and the e-folds (bottom) elapsed since the beginning of the simulation. The horizontal full line represents the (S$H_{0}$ES) energy density of the Universe at present. As shown, after it briefly becomes important near equality, the density of EDE scalar field is reducing drastically and it is orders of magnitude smaller than that of the matter background by the time of decoupling (\mbox{$z_{\rm dec}=1090$}), as required.}
\label{fig:densities}
\end{figure}

The barotropic parameters, of EDE and the background, are shown in
Fig.~\ref{fig:barotropic}. It can be seen clearly that, after thawing,
the barotropic parameter of EDE is \mbox{$w_\phi=1$} and the field is in
free-fall as discussed. Its density decreases as $a^{-6}$ as clearly shown
in Fig.~\ref{fig:densities}, which corresponds to the \mbox{$n\rightarrow\infty$}
limit of the oscillating EDE in Ref.~\cite{stringaxiverse} and it is never
attained by any oscillating EDE model. Thus, our model disturbs the emission of the CMB  at decoupling in
the least amount possible. 

Finally, for our example we obtain that the total excursion of the EDE field
from thawing to refreezing is sub-Planckian: \mbox{$\Delta\phi/\m=0.4274$}, in
agreement with Eq.~\eqref{Dphi}.
This implies both that our model does not suffer from fifth force problems and our
potential is stable against radiative corrections.

\section{Trapping at the origin}
\label{sec:ESP}

A compelling explanation why the EDE scalar field finds itself frozen at the
origin in the first place is the following. 
If the origin is an enhanced symmetry point (ESP), then at very early times, an
interaction of $\varphi$ with some other scalar field $\sigma$ can trap the
rolling $\varphi$ at zero~\cite{Kofman:2004yc}.
The scalar potential includes the interaction
\begin{equation}
    \Delta V=\frac12g^2\varphi^2\sigma^2\,,
    \label{DV}
\end{equation}
where the coupling $g<1$ parametrises the strength of the interaction.

We assume that initially $\varphi$ is rolling down its steep potential, which
away from the origin, does not have to be of the form in Eq.~\eqref{Vvarphi}.
In fact, it is conceivable that $\varphi$ might play the role of the inflaton
field too \cite{ours}. The original kinetic energy density of $\varphi$ is
depleted due to particle production of $\sigma$-particles, because their mass 
$\sim g\varphi$ changes non-adiabatically near the origin
\cite{Kofman:2004yc}. Note that, near the origin, 
the $\varphi$-field is approximately canonically normalised.

As the field moves past the ESP, the produced $\sigma$ particles give rise to
an effective linear potential
\mbox{$\sim gn_\sigma|\varphi|$} \cite{Kofman:2004yc}, where $n_\sigma$ is the
number density of the produced $\sigma$-particles.
This linear potential halts the roll of $\varphi$ and reverses its variation.
More $\sigma$-particles are created when $\varphi$ crosses the origin again,
resulting in a steeper linear potential, which reverses the variation of
$\varphi$ again, closer to the origin this time.
The process continues until the $\varphi$-field is trapped at the origin \cite{kostasbook,Kofman:2004yc}.

The trapping of a rolling scalar field at an ESP can take place only if the
$\sigma$-particles do not decay at maximum displacement.
The end result of this process is that all the kinetic energy density of the
rolling $\varphi$ has been given to the $\sigma$-particles. Since
$\varphi$ is trapped at zero, the $\sigma$-particles are relativistic, which
means that their density scales as radiation, being a subdominant part of the
thermal bath. 
As far as $\varphi$ is concerned, it is trapped at the origin and its density
is \mbox{$\rho_\varphi=V(\varphi=0)=e^{-\lambda} V_X=\,$constant}.

After some time, the $\sigma$-particles may decay into the standard model
particles, which comprise the thermal bath of the hot Big Bang. Because the
confining potential is proportional to $n_\sigma$, it disappears. However, the
EDE $\varphi$-field remains frozen at the origin because the scalar potential
$V(\varphi)$ in Eq.~\eqref{Vvarphi} is flat enough there. The EDE
$\varphi$-field unfreezes again in matter-radiation equality.


The above scenario is one possible explanation of the initial condition
considered. Numerical simulations simply assume that the field begins frozen at
the origin. Other possibilities to explain our initial condition exist, for
example considering a thermal correction of the form
\mbox{$\delta V\propto T^2 \varphi^2$}, which would drive the $\varphi$-field
towards the origin at high temperatures.

\section{Conclusions}

The concordance model $\Lambda$CDM suffers from the Hubble tension at 5$\sigma$.
A prominent resolution of this tension is early dark energy (EDE). EDE amounts
to a dark energy substance, which momentarily becomes about 10\% of the total
energy density near matter-radiation equality, but decays faster than radiation
afterwards.

EDE in the context of $\alpha$-attractors can unify EDE with late dark energy
without more fine-tuning than $\Lambda$CDM. We studied such a model of EDE,
characterised by the exp(exp) potential in Eq.~\eqref{Vvarphi}. Our EDE is
originally frozen at the origin. Near equality it thaws, then it free-falls
down its runaway potential until it refreezes before today, when it becomes 
late dark energy.

We have investigated numerically our model and demonstrated that it works for
natural values of the parameters. We also showed that the field excursion
between the initial and final frozen values is sub-Planckian, which means that
our model does not suffer from a fifth force problem and it is not unstable
against radiative corrections.


{\bf Acknowledgements:}
LB is supported by STFC.
KD is supported (in part) by the Lancaster-Manchester-Sheffield Consortium for Fundamental
Physics under STFC grant: ST/T001038/1.
SSL is supported by the FST of Lancaster University.
For the purpose of open access, the authors have applied a Creative Commons Attribution (CC BY) licence to any Author Accepted Manuscript version arising.

\end{document}